\documentclass[aps,prd,twocolumn,nofootinbib,superscriptaddress]{revtex4-2}

\usepackage[utf8]{inputenc}
\usepackage{bm}
\usepackage{amssymb}
\usepackage{amsmath}
\usepackage{graphicx}
\usepackage{tabularx}
\usepackage{enumitem}
\usepackage{color}
\usepackage{subcaption}
\usepackage{hyperref}
\usepackage{booktabs}
\usepackage{array}
\usepackage{upgreek}
\usepackage{appendix}
\usepackage[justification=raggedright,singlelinecheck=false,format=plain,labelsep=colon]{caption}

\makeatletter
\renewcommand\appendix{
	\par
	\setcounter{section}{0}%
	\setcounter{subsection}{0}%
	\setcounter{subsubsection}{0}%
	\setcounter{figure}{0}%
	\setcounter{table}{0}%
	\renewcommand{\thesection}{Appendix \Alph{section}}%
	\renewcommand{\thefigure}{\thesection.\arabic{figure}}%
	\renewcommand{\thetable}{\thesection.\arabic{table}}%
}
\makeatother

\hypersetup{
	colorlinks = true,
	linkcolor = blue,
	citecolor = blue,
	filecolor = blue,
	urlcolor = blue
}

\def\apj{ApJ}
\def\apjl{ApJL}
\def\apjs{Astrophys. J. Supp.}
\def\araa{Ann. Rev. Astron. Astrophys.}
\def\aap{Astron. Astrophys.}

\def\mnras{MNRAS}
\def\nat{Nature}

\def\physrep{Phys. Rep. }

\newcolumntype{K}[1]{>{\centering\arraybackslash}p{#1}}

\let\OLDthebibliography\thebibliography
\renewcommand\thebibliography[1]{
	\OLDthebibliography{#1}
	\setlength{\parskip}{0pt}
	\setlength{\itemsep}{0pt plus 0.3ex}
}

\allowdisplaybreaks

\begin{document}
	
	\title{Self-gravitating isothermal sphere in an expanding background}
	
	\author{Zacharias Roupas}
%	\email{Zacharias.Roupas@unimib.it}
	\affiliation{Dipartimento di Fisica ``G. Occhialini'', 
		Universit\'a degli Studi di Milano-Bicocca, Piazza della Scienza 3, 20126 Milano, Italy}
	\affiliation{Istituto Nazionale di Fisica Nucleare (INFN), Sezione di Milano-Bicocca, 
		Piazza della Scienza 3, 20126 Milano, Italy}
		
%	\date{}
	
	\begin{abstract}
		Spatially homogeneous thermal equilibria of self-gravitating gas, being impossible otherwise,  are nevertheless allowed in an expanding background accounting for Universe's expansion. 
		Furthermore, a fixed density at the boundary of a perturbation is a natural boundary condition keeping the mass finite inside without the need to invoke any unphysical walls.
		These facts allow us to develop a consistent gravitational thermodynamics of isothermal spheres inside an expanding Universe.
		In the canonical and grand canonical ensembles we identify an instability for both homogeneous and inhomogeneous equilibria.
		We discuss a potential astrophysical application.
		If such an instability is triggered on baryonic gas at high redshift $z > 137$ when the primary baryonic component, namely atomic hydrogen, was still thermally locked to the Cosmic Microwave Background radiation, then the corresponding destabilized gaseous clouds have baryonic mass $\geq 0.8\cdot 10^5 {\rm M}_\odot$ and radius $\geq 15{\rm pc}$.
	\end{abstract}
	
	\maketitle
	
	\section{Introduction}
	
	In the standard, $\Lambda$CDM, model of Cosmology space is expanding within the framework of General Relativity \citep{1993ppc..book.....P}. In the weak field limit approximation and in comoving coordinates the expansion introduces an apparent inertial force, allowing for homogeneous equilibria.
	We develop the thermodynamics of self-gravitating gas in this Newtonian limit. We consider an isothermal sphere at some radius, negligible with respect to Hubble radius, and develop the stability analysis for canonical and grand canonical perturbations and fixed boundary density, accounting for the apparent inertial force of space expansion.
	
	The statistical ensembles are not equivalent for self-gravitating systems, regarding their stability properties, because of gravity's non-additivity \citep{1990PhR...188..285P,2003FoPh...33..223K}. We show that the boundary conditions of a perturbation characterize the respective ensemble and are in effect responsible for their non-equivalence. Therefore, the same gravitational equilibrium at different boundary conditions may present different stability properties, and we specify these for our case.
	
	Besides the standard homogeneous state in an expanding Universe, we discover additional inhomogeneous equilibria and study their stability as well. We discover that both series of equilibria undergo instabilities in both ensembles. We further discuss a potential astrophysics application of our results. 
	At the early stages of Universe's evolution, during the dark ages, baryonic matter is in the form primarily of neutral atomic hydrogen and its temperature is coupled with the Cosmic Microwave Background (CMB) radiation for
	 $1100 \gtrsim z \gtrsim 137$, 
	 because of a small electron density, remnant from the recombination era \citep{1993ppc..book.....P,1999ApJ...523L...1S,2000ApJS..128..407S,2001PhR...349..125B,2020A&A...641A...6P}. Therefore this monotatomic Hydrogen gas is isothermal and during this era matter is homogeneously distributed with 
	 $\delta \rho/\rho \lesssim 10^{-5}$ \citep{1993ppc..book.....P}. These conditions are consistent with our framework.
	Recent observational data, primarily from the James Webb Space Telescope (JWST), have revealed a striking overabundance of luminous, mature, galaxies at high redshift even at $z>10$
	\citep{2016ApJ...824...21S,2024arXiv240803920P,2023MNRAS.518.4755A,2023MNRAS.519.1201A,2022ApJ...938L..15C,2023MNRAS.518.6011D,2023Natur.616..266L,2022ApJ...940L..55F,2023ApJ...946L..35M,2023ApJ...942L...9Y,2022ApJ...940L..14N}.
	These observations suggest a strong tension of $\Lambda$CDM with high-redshift galaxies observations
	\citep{2023NatAs...7..731B}. 
	In addition, JWST revealed an overabundance of Active Galactic Nuclei (AGN) in the first billion years of Universe's life that appear already mature in their evolution ($z\gtrsim 6$)
	\citep{2024ApJ...964...39G,2024Natur.627...59M,2024ApJ...963..129M,2023ApJ...959...39H,2023arXiv230801230M,2024NatAs...8.1054B}. These Massive Black Holes (MBH) appear to be also impossibly massive posing a key challenge to theoretical models of MBH formation (e.g. \citep{2023ARA&A..61..373F,2020ARA&A..58...27I}), most probably linked with the early galaxies problem.
	These serious challenges at high redshift may indicate that non-linear small-scale structure formation have begun earlier and/or progressed faster than predicted within the standard hierarchical merging paradigm. The thermodynamic instability identified in this work supports these ideas.
	
	In the next section we review the Poisson equation in comoving coordinates in an expanding background. In section \ref{sec:equilibria} we calculate the thermodynamic equilibria, and in section \ref{sec:stability} we perform the thermodynamic stability analysis by calculating explicitly unstable modes.
	We discuss the potential astrophysical application in section \ref{sec:astrophysics}.

	\section{Poisson equation in canonical comoving coordinates}\label{sec:poisson}
	
	In the weak field limit, the time-time component of the Einstein field equations give for an ideal fluid (e.g. \citep{1972gcpa.book.....W})
	\begin{equation}\label{eq:poisson_proper}
		\nabla_r^2\Phi = 4\pi G \left(\rho + 3\frac{P}{c^2} \right) - \Lambda c^2
	\end{equation}
	where $\rho$ denotes the mass density of the fluid, $P$ the pressure, and $\Lambda$ the cosmological constant. 
	The $\rho$ and $P$ may deviate locally from the homogeneous, at large scales, background. Therefore, a pressure gradient $\nabla P \neq 0$ may develop even for non-relativistic, so-called in cosmological context ``pressureless'', matter $P/c^2 \ll \rho $.
	The differentiation is with respect to the proper Minkowski coordinates $\bm{r}(t)$. These are related to the comoving coordinates $\bm{x}$ by
	\begin{equation}\label{eq:coordinates}
		\bm{r} = a\cdot \bm{x},
	\end{equation}
	where $a(t)$ is the scale factor that describes the expansion of the Universe
	\begin{align}
		\label{eq:expansion_kin}
		&\left(\frac{\dot{a}}{a}\right)^2 = \frac{8\pi G}{3}\rho_{\rm b} + \frac{\Lambda c^2}{3} - \frac{kc^2}{a^2} \\
		\label{eq:expansion_acc}
		&\frac{\ddot{a}}{a} = -\frac{4\pi G }{3}\left(\rho_{\rm b} + \frac{3P_{\rm b}}{c^2} \right) + \frac{\Lambda c^2}{3}.
	\end{align}
	The subscript `$b$' denotes that $\rho_{\rm b} = \rho_{\rm b} (t)$, $P_{\rm b} = P_{\rm b} (t)$ are the background constant density and pressure, and $k = 0,\pm 1$ determines the 3-geometry of space.
	
	The Lagrangian of a particle with mass $m$ in this expanding background in the proper coordinates is
	\begin{equation}\label{eq:L_proper}
		\mathcal{L}(r,t) = \frac{1}{2}m\dot{r}^2 - m\Phi.
	\end{equation}
	We are interested in deviations from the background and therefore we assume that the proper velocity 
	\begin{equation}\label{eq:rdot}
		\dot{\bm{r}} = \dot{a}\bm{x} + a\dot{\bm{x}}
	\end{equation}
	involves not only the velocity of expanding space $\dot{a}\bm{x}$ but also a non-negligible \textit{motion relative to the expanding background} 
	\begin{equation}\label{eq:v} 
		\bm{v} = a\dot{\bm{x}}.
	\end{equation}
	This is the proper peculiar velocity, that is the one measured by an observer at the particle position and fixed $\bm{x}$. We note that it varies within thermal relaxation timescales, while $\dot{a}$ varies within the much vaster cosmic timescales. Temperature is related to $\left\langle \bm{v}^2 \right\rangle$.
	
	As described by Peebles \citep{1980lssu.book.....P}, the transformation from the proper locally Minkowski coordinates $\bm{r}$ to comoving coordinates $\bm{x} = \bm{r}/a$ leads to the equivalent Lagrangian
	\begin{equation}\label{eq:L_comoving_tr}
		L(\bm{x},t) = \mathcal{L}(\bm{r},t) + \frac{d F}{d t}, \;
		F = \frac{1}{2}m a \dot{a} x^2,
	\end{equation}
	that is
	\begin{equation}\label{eq:L_comoving}
		L = \frac{1}{2}m(a\dot{\bm{x}})^2 - m\phi 
	\end{equation}
	where the new potential is 
	\begin{equation}\label{eq:phi}
		\phi = \Phi + \frac{1}{2} a \ddot{a} x^2.
	\end{equation}
	It is straightforward to verify that $L$ and $\mathcal{L}$ give the same equations of motion.
	
	The proper and comoving momenta are, respectively, 
	\begin{align}
		\label{eq:p_r}
		&\bm{p_r} \equiv \frac{\partial \mathcal{L}}{\partial \dot{\bm{r}}} = m\dot{\bm{r}},\\
		\label{eq:p}
		&\bm{p}  \equiv \frac{\partial L}{\partial \dot{\bm{x}}} = ma^2\dot{\bm{x}},
	\end{align}
	and the proper Hamiltonian
	\begin{equation}\label{eq:H_proper}
		\mathcal{H}(\bm{r},\bm{p_r},t) \equiv \dot{\bm{r}}\bm{p_r} - \mathcal{L} = \frac{\bm{p_r}^2}{2m} + m\Phi(\bm{r},t).
	\end{equation}
	Therefore, the transformations
	\begin{align}
		\label{eq:canonical_coo}
		\bm{x}(\bm{r},\bm{p_r}) &= \frac{1}{a}\bm{r},\quad 
		\bm{p}(\bm{r},\bm{p_r}) = \frac{1}{a}\bm{p_r} - m\frac{\dot{a}}{a}\bm{x},
		\\
		\label{eq:canonical_H}
		H(\bm{x},\bm{p},t) &\equiv  	\mathcal{H}(\bm{r},\bm{p_r},t) - \bm{r}\bm{p_r} + \bm{x}\bm{p} - \frac{dF}{dt} 
		\nonumber \\
		&= \frac{\bm{p}^2}{2m} + m\phi (\bm{x},t)
	\end{align}
	are canonical.
	
	The Poisson equation for $\phi$ in the comoving coordinates becomes
	\begin{equation}\label{eq:poisson_comoving_P}
		\nabla^2 \phi = 4\pi G a^2 \left((\rho - \rho_{\rm b}) + 3\frac{P-P_{\rm b}}{c^2} \right)
	\end{equation}
	where the differentiation now is with respect to the comoving coordinates $\bm{x}$. The cosmological constant cancels out in comoving coordinates \citep{1980lssu.book.....P}. We shall consider here non-relativistic matter 
	\begin{equation}
		P_{\rm m} \ll \rho_{\rm m} c^2
	\end{equation}
	with 
	$\rho = \rho_{\rm m} + \rho_{\rm b}$, $P = P_{\rm m} + P_{\rm b}$ .
	We shall consider perturbation only in the baryonic component and not in the radiation component which we consider to remain homogeneous, not affected by baryonic perturbations.
 	Therefore the relativistic background pressure and the radiation pressure cancel out in (\ref{eq:poisson_comoving_P})
	\begin{equation}\label{eq:poisson_comoving}
		\nabla^2 \phi = 4\pi G a^2 \left(\rho - \rho_{\rm b}\right)
	\end{equation}
	
	Equation (\ref{eq:poisson_comoving}) identifies the source for $\phi$ as the fluctuating part with respect to a constant density equilibrium state, so that there is no need to invoke the Jeans swindle in an expanding Universe. It has been overlooked nevertheless that it allows for inhomogeneous equilibria, static in comoving coordinates, to develop. Most importantly for our purposes, defining equilibria --both homogeneous and inhomogeneous-- static in comoving coordinates within a certain volume by eq. (\ref{eq:poisson_comoving}) allows for the rigorous theoretical development of thermodynamics in expanding background. 
	
	We consider a finite region of space with radius $R$ (equivalently $X$ in the proper, comoving coordinates)
	\begin{equation}\label{eq:RX}
		R=aX,
	\end{equation}
	with $R$ much lower than the Hubble horizon. Then equation (\ref{eq:poisson_comoving}) describes the distribution of matter under its self-gravity through a Newtonian self-gravitating potential $\phi_{\rm N}$ and additionally under the influence of an external potential $\phi_{\rm ext} = \frac{1}{2} \ddot{a} a x^2$, which generates in comoving coordinates the apparent inertial force due to Universe expansion. The matter density is decomposed as
	\begin{equation}
		\rho = m n + \rho_{\rm DM},
	\end{equation}
	where $n$ is the gas number density, $m$ the mass of one baryon, and $\rho_{\rm DM}$ denotes the dark matter density.
	We assume here that
	\begin{equation}\label{eq:DM_assumption}
		\rho_{\rm DM} = \rho_{\rm DM, b} = {\rm const}
	\end{equation}
	within $R$. 
	Equation (\ref{eq:DM_assumption}) preassumes that dark matter content within $R$ shall not be affected by gas perturbations of scale $R$.
	Therefore the dark matter contribution in (\ref{eq:poisson_comoving}) cancels out. Thus, we get under the assumption (\ref{eq:DM_assumption})
	\begin{equation}\label{eq:phi_sum}
		\phi = \phi_{\rm N} + \phi_{\rm ext},
	\end{equation}
	with
	\begin{eqnarray}
		\label{eq:phi_N_P}
		\phi_{\rm N}(x) &=& - Ga^2 m \int^{\{X\}}\frac{n(\bm {x}) }{\left|\bm{x} - \bm{x}^\prime \right|} d^3\bm{x}^\prime,
		\\
		\label{eq:phi_ext_P}
		\phi_{\rm ext}(x) &=&  -\frac{2\pi G}{3}a^2 m n_{\rm b} x^2 .
	\end{eqnarray}
	Here, and in the followings, we suppress the dependence on cosmic time $t$ considering equilibria and (static) perturbations within thermal relaxation timescales which are much smaller than the cosmic timescales $t$, in the astrophysics application we discuss (see \ref{app:t_rel}). Therefore $n$, $a$, which vary at cosmic timescales, are constant in thermal relaxation timescales. The dependence on time within thermal relaxation timescales is expressed by the momenta $p$.
	
	\section{Thermodynamic equilibria}\label{sec:equilibria}
	
	We consider atomic Hydrogen contained within some volume
	\begin{equation}\label{eq:V}
		V = \frac{4}{3}\pi R^3 = \frac{4}{3}\pi a^3 X^3.
	\end{equation}
	Assuming spherical symmetry, the number of particles $\nu(\bm{x})$ (we reserve $N$ to denote the total number of particles in $R$) at each point $\bm{x}$ within $R = a\, X$ are given by
	\begin{equation}\label{eq:Nx}
		\frac{d\nu}{dx} = 4\pi a^3 m\, n(x)\, x^2 .
	\end{equation} 
	The gas within $R$ is subject to the boundary condition
	\begin{equation}\label{eq:bc}
		n_R \equiv n(X) = n_{\rm b} = {\rm const.},
	\end{equation} 
	As we already noted, we further assume that the thermal relaxation timescale of the monoatomic gas within $R$ is significantly smaller than the gravitational dynamical timescale, as we show in \ref{app:t_rel}, so that the gas thermalizes rapidly with respect to any gravitational perturbations. The background serves therefore as a heat bath of constant temperature $T$ (adiabatic invariant at cosmic timescales) and the gas within $R$ attains a thermal equilibrium at this temperature.
	
	The gas is subject to the Poisson equation 
	\begin{equation}\label{eq:poisson_comoving_n}
		\nabla^2\phi(\bm{x}) = 4\pi G m a^2 \left(n(\bm{x}) - n_{\rm b} \right),
	\end{equation}
	with $\phi = \phi_{\rm N} + \phi_{\rm ext}$, where the respective self-gravity Newtonian and external potentials are given in (\ref{eq:phi_N_P}), (\ref{eq:phi_ext_P}).
	
	The comoving coordinates with the natural boundary condition (\ref{eq:bc}) and the external potential (\ref{eq:phi_ext_P}) allows performing gravitational thermodynamics within $R$ without invoking any unphysical ``walls'' to maintain a finite mass within this volume. There is a finite boundary pressure exerted on the gas by the background environment, which also serves as a heat bath. 
	
	We introduce the non-dimensional volume element
	\begin{equation}
		d\bm{\tau} \equiv \frac{d^3\bm{p}}{h^3}d^3\bm{x}.
	\end{equation} 
	The constant $h$ has dimensions of action. It represents the elementary phase-space volume and in our application $h$ may be identified with Planck's constant. We shall consider the one-particle distribution function  $f(\bm{x},\bm{p})$, that is appropriate in the mean-field limit \citep{1990PhR...188..285P,2003FoPh...33..223K}, so that
	\begin{equation}\label{eq:Ntot_def}
		N = \iint_{\{ V \}} f(\bm{x},\bm{p}) d\bm{\tau},
	\end{equation}
	where $N \equiv \nu(X)$ the total number of particles within $R = a X$.
	The corresponding Boltzmann entropy is 
	\begin{equation}\label{eq:S_def}
		S = - \iint_{\{ V \}} f(\bm{x},\bm{p}) \ln f(\bm{x},\bm{p}) d\bm{\tau}.
	\end{equation}
	The number density at each point $\bm{x}$ is derived from the distribution function from the integral
	\begin{equation}\label{eq:n_def}
		n(\bm{x}) = a^{-3} \int f(\bm{x},\bm{p}) \frac{d^3\bm{p}}{h^3}.
	\end{equation}
	The total energy is given as
	\begin{equation}\label{eq:E_def}
		E = \iint_{\{ V \}} \frac{\bm{p}^2}{2m} f(\bm{x},\bm{p}) d\bm{\tau} + U,
	\end{equation}
	where the potential energy is
	\begin{align}\label{eq:U_def}
		U = &-\frac{1}{2}\frac{Gm^2}{a}\iiiint_{\{ V \}} \frac{f(\bm{x},\bm{p}) f(\bm{x}^\prime,\bm{p}^\prime) }{\left|\bm{x} - \bm{x}^\prime \right|} d\bm{\tau} d\bm{\tau^\prime} +
		\nonumber \\
		&+ m\iint_{\{ V \}} \phi_{\rm ext}(\bm{x}) f(\bm{x},\bm{p}) d\bm{\tau}.
	\end{align}
	We consider $S=S[f]$, $N =N [f]$, $E=E[f]$ as functionals of $f$ and consider their first-order variations $\delta^{(1)} S \equiv \frac{\partial S}{\partial f}\delta f$ under small perturbations $\delta f$. 
	Equivalently, we consider perturbations in $f$
	\begin{equation}
		f_{\rm pe} = f + \delta f,
	\end{equation}
	where the unscripted $f$ denotes the equilibrium.
	The thermodynamics quantities' variation include $\delta f$ at several orders, e.g. 
	\begin{equation} 
		\delta S = \delta^{(1)}S + \delta^{(2)}S + \cdots,
	\end{equation}	
	where
	\begin{equation} 
	\delta^{(1)}S = \mathcal{O}(\delta f),\;
	\delta^{(2)}S = \mathcal{O}(\delta f^2),\; \ldots
	. 
	\end{equation}	
		
	The equilibrium configurations are stationary points of the entropy for fixed energy, mass and volume
	\begin{equation}\label{eq:1st_order}
		\delta^{(1)} S|_V - \beta \delta^{(1)} E |_V + \alpha \delta^{(1)} N |_V = 0 
	\end{equation}
	where $\beta$ and $\alpha$ are Lagrange multipliers. 
	
	Denoting $f^\prime \equiv f(\bm{x}^\prime,\bm{p}^\prime)$, the variation of potential energy is 
	\begin{align}\label{eq:dU1}
		\delta^{(1)} U = &-\frac{1}{2}\frac{Gm^2}{a}\iiiint_{\{ V \}} \frac{f^\prime (\delta f) + f (\delta f^\prime) }{\left|\bm{x} - \bm{x}^\prime \right|} d\bm{\tau} d\bm{\tau^\prime} + 
		\nonumber \\
		&+ m\iint_{\{ V \}} \phi_{\rm ext} (\delta f) d\bm{\tau}
		= m\iint_{\{ V \}} \phi (\delta f) d\bm{\tau}, 
	\end{align} 
	Consequently, the first-order variation (\ref{eq:1st_order}) gives
	\begin{align}\label{eq:dS1_int}
		&\iint_{\{V\}} (\delta f)\,d\bm{\tau } \times
		\nonumber \\
		&\left\lbrace -(1+\ln f(\bm{x},\bm{p}) ) - \beta \frac{\bm{p}^2}{2m} - \beta m\phi(\bm{x}) + \alpha \right\rbrace = 0,
	\end{align}
	and therefore the equilibrium phase-space distribution is
	\begin{equation}\label{eq:feq_alpha}
		f(x,p) = e^{\alpha -1}e^{-\beta\frac{p^2}{2m}} e^{-\beta m\phi(x)}.
	\end{equation}
	The normalization condition (\ref{eq:n_def}) gives
	\begin{equation}\label{eq:feq_n}
		f(x,p) = \frac{n(x)}{n_h}e^{-\beta\frac{p^2}{2m}} 
		,\quad 
		n_h \equiv \left(\frac{\beta }{2\pi m} \right)^{3/2} a^3 h^3,
	\end{equation}
	and the number density can be written as
	\begin{equation}\label{eq:n_bol}
		n(x) = n_{\rm b} e^{-m\beta(\phi(x)-\phi(R))},
	\end{equation}
	with $n_0 \equiv n(0)$. 
	
	Therefore, $\beta$ does indeed correspond to the inverse temperature related to the kinetic energy as
	\begin{equation}\label{eq:K}
		K \equiv \iint_{\{ V \}} \frac{\bm{p}^2}{2m} f(\bm{x},\bm{p}) d\bm{\tau} = \frac{3}{2}\frac{N}{\beta}.
	\end{equation}
	In \ref{app:hydrostatic} we show the equivalence of our thermodynamic equilibria with hydrostatic equilibria for an ideal gas equation of state. 
	
	The Lagrange multiplier $\alpha$ (not to be confused with the cosmological scale factor $a$) defines the chemical potential $\mu \equiv \alpha /\beta$. It depends on the boundary potential as
	\begin{equation}\label{eq:alpha}
		\alpha = 1 + \beta m\phi(R) + \ln \frac{n_{\rm b}}{n_h} .
	\end{equation}
	The boundary potential may be calculated by direct integration of (\ref{eq:phi_N_P}) using also (\ref{eq:Ntot_def})
	\begin{equation}\label{eq:phiR} 
		\phi(R) = - G \frac{m N}{R} - \frac{2\pi G}{3} m n_{\rm b} R^2 . 
	\end{equation}
	
	Equation (\ref{eq:n_bol}) relates the potential with the density and therefore allows us to integrate Poisson equation (\ref{eq:poisson_comoving_n}) and identify the inhomogeneous equilibria $n(x)$, besides the homogeneous solution $n = n_{\rm b}$. 
	
	Since we consider fixed boundary density $n_{\rm b}$ and temperature $\beta^{-1}$, the Jeans length 
	\begin{equation}\label{eq:R_J}
		R_{\rm J} \equiv \sqrt{\frac{\pi}{Gm^2n_{\rm b}\beta}},\quad X_{\rm J} \equiv \frac{R_{\rm J}}{a}.
	\end{equation}
	introduces a natural length scale that allows us to define non-dimensional quantities that implicitly manifest our constraints. In particular we define 
	\begin{equation}\label{eq:nond_def}
		N_{\rm J} \equiv n_{\rm b} \frac{4}{3}\pi \left(\frac{R_{\rm J}}{2}\right)^3,\;
		\tilde{x} \equiv  \frac{x}{X_{\rm J}},\;
		{\zeta} \equiv  \frac{R}{R_{\rm J}},\;
		\tilde{\nu} = \frac{\nu}{6 N_{\rm J}},
	\end{equation}
	and
	\begin{equation}\label{eq:y_def}
		y \equiv \beta m (\phi(x) - \phi(R)) .
	\end{equation}
	It shall be clear that the ``Jeans'' length and mass used here are only meant as characteristic scales and do not correspond to the radius and mass values of the onset of the thermodynamic instability. These shall be calculated in the next section. We emphasize, nevertheless, that the Jeans mass $mN_{\rm J}$, as above, remains constant during the whole cosmic time interval $1100 \lesssim z \lesssim 137$ (e.g. \citep{2001PhR...349..125B}), when $n_{\rm b} \propto (1+z)^3$ and $\beta^{-1} \propto (1+z)$ because the gas remains thermally coupled with the CMB radiation.
	We shall see that the instability's critical mass shall be proportional to the Jeans mass and therefore shall also be constant throughout the dark ages timeline we consider in the Astrophysics application, section \ref{sec:astrophysics}.
	
\begin{figure}[tb]
		\centering
		\includegraphics[width=0.9\columnwidth]{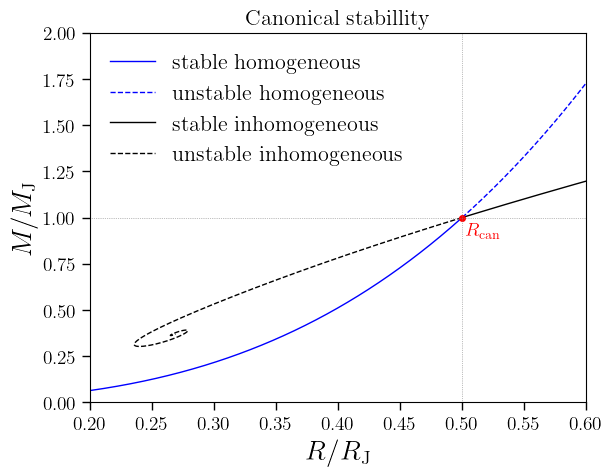}  
		\caption{Series of thermodynamic equilibria, expressed by total mass $M(R)$ contained within radius $R$, of self-gravitating monoatomic gas in an expanding background in the canonical ensemble.}
		\label{fig:MR_series_can}
	\end{figure}

	Equations (\ref{eq:Nx}), \ref{eq:poisson_comoving_n}), (\ref{eq:n_bol}) define the system of equations in non-dimensional variables 
	\begin{equation}
		\label{eq:equilibrium}
		\frac{dy}{d\tilde{x}} = \frac{4\pi^2}{3}\left(\frac{\tilde{\nu}}{\tilde{x}^2} - \tilde{x} \right)
		\, ,\quad
		\frac{d\tilde{\nu}}{d\tilde{x}} = 3 \tilde{x}^2 e^{-y},
	\end{equation}
	whose solution for $\tilde{x} \in {[0,\zeta]}$ specifies the homogeneous equilibrium and the inhomogeneous equilibria under the boundary conditions
	\begin{equation}\label{eq:bcs}
		y(0) = - \ln{ \frac{n_0}{n_{\rm b}} },\;
		y(\zeta ) = 0,\;
		\tilde{\nu}(0) = 0 .
	\end{equation}
	We integrate this system directly for several radii $\zeta = R/R_{\rm J}$ inwards to get a series of inhomogeneous equilibria. Exactly at $R = \frac{1}{2}R_{\rm J}$ the inhomogeneous series crosses the homogeneous series. We depict the series of equilibria in Figures \ref{fig:MR_series}, \ref{fig:MR_series_can}, discussed in sections \ref{sec:stability_gc}, \ref{sec:stability_can}.
	
	\section{Stability analysis}\label{sec:stability}
	
	\subsection{Grand canonical perturbations}\label{sec:stability_gc}
	
	The second law of thermodynamics bounds statistically the amount of heat $\Delta Q$ that may be exchanged between the system and its environment for any processes, reversible or irreversible, as
	\begin{equation}\label{eq:2nd_law}
		\Delta Q \leq T \Delta S.
	\end{equation}
	This inequality defines the spontaneous, preferable direction of evolution of the system.
	For an isolated system of fixed volume, with no exchange of heat, the entropy is therefore statistically increasing $\Delta S \geq 0$. This suggests that the entropy attains its maximum possible value at a stable equilibrium. Therefore the condition for a stable equilibrium for isolated systems is that the entropy variation for a perturbation about the equilibrium is negative $\delta^{(2)}S < 0$, i.e. the stable equilibrium is an entropy maximum. 
		
	This reasoning is generalized for open systems utilizing the inequality (\ref{eq:2nd_law}) and the first law of thermodynamics, which for fixed volume is
		\begin{equation}\label{eq:1st_law}
		\Delta E = \Delta Q  + \mu \Delta  N .
	\end{equation}
	We get that the system is evolving such that the following quantity is decreasing
	\begin{equation}
		\Delta J \equiv	\beta \Delta E  - \Delta S - \alpha \Delta  N \leq 0,
	\end{equation}
	where $\alpha = \mu \beta$, $\beta = T^{-1}$.
	Thus, $J$ should be a minimum, $\delta^{(2)} J >0$, for any perturbation about a stable equilibrium.
	Therefore, the first-order variation (\ref{eq:1st_order}), equivalent to $\delta^{(1)} J = 0$, defines the thermodynamic equilibrium configurations and the second-order variation determines the statistical, thermodynamic stability of these configurations. An equilibrium is unstable if there exists any perturbation about it for which
	\begin{equation}\label{eq:2nd_order}
		\text{instability condition:} \quad \beta \delta^{(2)} E  - \delta^{(2)} S - \alpha \delta
		^{(2)} N < 0.
	\end{equation}
	The instability sets in exactly where
	\begin{equation}\label{eq:onset}
		\text{instability onset:} \quad \beta \delta^{(2)} E  - \delta^{(2)} S - \alpha \delta
		^{(2)} N = 0,
	\end{equation}
	that is where $\delta^{(2)} J$ changes sign. 
	
			\begin{table}[tb]
		\begin{center}
			\begin{tabular}{c  c || c c c}			
				$R/R_{\rm J}$
				&				
				$M/M_{\rm J}$
				&			
				$\lambda_0$
				&			
				$\lambda_1$
				&			
				$\lambda_2$
				\\
				\toprule
				$0.2742$
				&				
				$0.1238$
				&				
				$0$
				&			
				\color{green}{$-382.323$}
				&			
				\color{green}{$-1127.26$}
				\\
				\midrule
				$0.6212$
				&				
				$1.4387$
				&				
				$9.44562$
				&			
				$0$
				&			
				\color{green}{$-27.6689$}
				\\
				\midrule
				$0.9556$
				&				
				$5.2358$
				&
				$2.97406 $
				&			
				$5.17068 $
				&			
				$0$
			\end{tabular}
			\caption{The first three turning points of stability for grand-canonical perturbations, expressed by the respective radius $R$ and total mass $M$, for homogeneous equilibria and the corresponding first three eigenvalues. At each turning point one additional mode becomes unstable, i.e. one additional eigenvalue becomes zero. Green color denotes stable modes.}
			\label{tab:hom_eig}
		\end{center}
	\end{table}
	
	\begin{figure}[tb]
		\centering  
		\begin{subfigure}{\columnwidth}
			\centering  
			\includegraphics[width=0.85\textwidth]{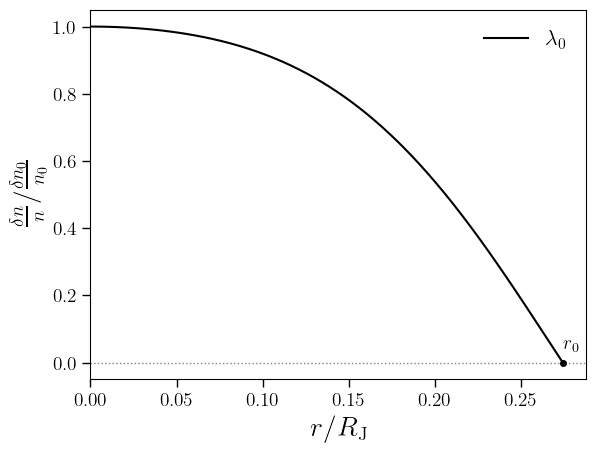} 
			\captionsetup{justification=centering} 
			\caption{$R = 0.2742 R_{\rm J}$}
			\label{fig:dn_R1_hom}
		\end{subfigure}
		\begin{subfigure}{\columnwidth}
			\centering 
			\includegraphics[width=0.85\textwidth]{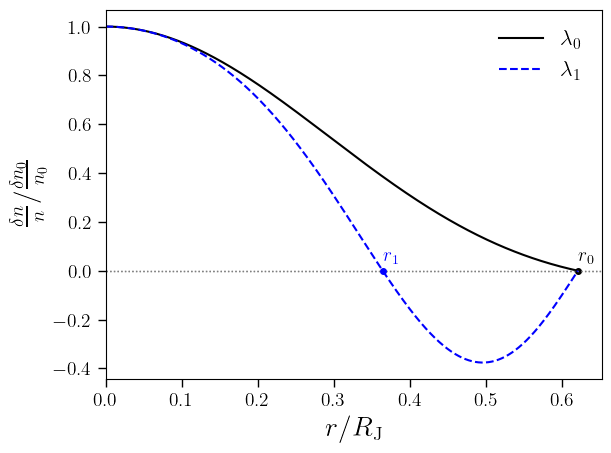} 
			\captionsetup{justification=centering}
			\caption{$R = 0.6212 R_{\rm J}$}
			\label{fig:dn_R2_hom}
		\end{subfigure}
		\begin{subfigure}{\columnwidth}
			\centering 
			\includegraphics[width=0.85\textwidth]{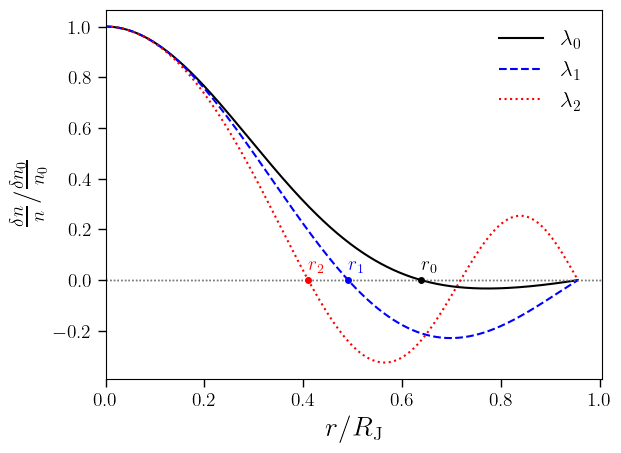} 
			\captionsetup{justification=centering}
			\caption{$R = 0.9556 R_{\rm J}$}
			\label{fig:dn_R3_hom}
		\end{subfigure}
		\caption{The eigenfunctions of the unstable modes of homogeneous equilibria for the first three turning points of stability, for grand canonical perturbations, corresponding to the eigenvalues of Table \ref{tab:hom_eig}.}
		\label{fig:dn_R_hom}
	\end{figure}

	Let us calculate in detail the second-order variation (\ref{eq:2nd_order}) and seek for potential turning points of stability.
	We consider perturbations only in the spatial part of the distribution function
	\begin{equation}
		n_{\rm pe} = n + \delta n,
	\end{equation}
	where $n$ denotes any equilibrium configuration either homogeneous or inhomogeneous. Perturbations in $n$ induce perturbations in the gravitational potential $\phi$ via
	\begin{equation}\label{eq:dphi}
		\frac{d}{dx}\delta \phi  = G\frac{m}{a x^2}\delta\nu,
	\end{equation}
	and recall that
	\begin{equation}\label{eq:dn}
		\delta n = \frac{1}{4\pi a^2 x^2}\frac{d}{dx}\delta \nu .
	\end{equation}
	Since $N = a^3 \int^{\{X\}} n d^3\bm{x}$, the second-order perturbation of the total mass with respect to density for fixed volume is zero
	\begin{equation}
		\delta^{(2)} N = 0,
	\end{equation}
	but still, for an open system and a perturbation about the equilibrium it is
	$\delta^{(1)}N \neq 0$ and therefore $\delta N \neq 0$, 
	affecting the stability as we will see through a second order term of the form $(\delta N)^2$.

	It is straightforward to calculate the second-order variation of entropy and energy (see \ref{app:inst_condition}). After some algebra we get
	\begin{eqnarray}
		\label{eq:dS2}
		\delta^{(2)}S &=& -\frac{1}{2} \int^{\{X\}} \frac{(\delta n)^2}{n} a^3 d^3 \bm{x}
		\\
		\label{eq:dE2}
		\delta^{(2)}E &=& \frac{1}{2} \int^{\{X\}} m (\delta n) (\delta \phi) a^3 d^3 \bm{x} .
	\end{eqnarray}
	We substitute (\ref{eq:dphi}), (\ref{eq:dn}), (\ref{eq:dS2}), (\ref{eq:dE2}) into (\ref{eq:2nd_order}) and after some integrations by parts with boundary conditions
	\begin{equation}
		\delta \nu(0) = 0,\quad
		\delta n (r=R) = 0, \quad
		\delta \nu (r=R) = \delta N, 
	\end{equation}
	where recall that $r=a x$, we get (see \ref{app:inst_condition}) the condition for \textit{instability}
	\begin{align}\label{eq:2nd_order_inst}	
		&-\frac{Gm^2 \beta}{2}\left\lbrace \frac{(\delta N)^2}{R} 
		+ \int_0^{R/a} \delta \nu(x) \times
		\right.
		\nonumber \\
		&\left. 
		\left(
		\frac{1}{a x^2} +
		\frac{d}{dx}\left(\frac{1}{4\pi G m^2 \beta a^3 x^2 n(x)}\frac{d}{dx} \right)
		\right)
		\delta \nu(x)
		dx
		\right\rbrace
		< 0 .
	\end{align}
	The density $n(x)$ denotes the equilibrium configuration about which we perturb the system. For the homogeneous equilibria it is $n = n_{\rm b} = {\rm const}$. 
	
		Consider the eigenvalues $\lambda$ 
	\begin{equation}\label{eq:K_eig}
		\int_0^{R/a} \hat{K}(x,x') q(x') dx' = \lambda q(x)
	\end{equation}
	of the kernel (for the use of a somewhat similar Kernel in the case of preserving total mass without an expanding background see \cite{1990PhR...188..285P})
	\begin{equation}
		\hat{K}(x,x') \equiv \updelta (x - \tfrac{R}{a}) \updelta (x' - \tfrac{R}{a}) \frac{1}{a x'} 
		+ 
		\updelta(x-x') \hat{T}(x'),
	\end{equation}
	where $\updelta$ denotes the Dirac delta and
	\begin{equation}
		\hat{T}(x) \equiv \frac{1}{a x^2}
		+
		\frac{d}{dx}\left(\frac{1}{4\pi G m^2 \beta a^3 x^2 n(x)}\frac{d}{dx} \right) .
	\end{equation}
	Multiplying (\ref{eq:K_eig}) by $q(x)$ and integrating over $x$ we get
	\begin{align}
		&\int_0^{R/a} dx\int_0^{R/a} q(x) \hat{K}(x,x') q(x') dx' = \lambda \int_0^{R/a}q(x)^2 dx
		\nonumber \\
		\label{eq:eig_lx}
		&\frac{1}{R}q(\tfrac{R}{a})^2 + 
		\int_0^{R/a} q(x)\hat{T}(x) q(x) dx = \lambda \int_0^{R/a}q(x)^2 dx.
	\end{align}
	Comparing this with the condition for instability (\ref{eq:2nd_order_inst}) it is clear that the sign of the eigenvalues $\lambda$ determine the stability of the system, with $\lambda >0$ signifying an unstable mode, as long as we identify $q = \delta \nu$ with
		\begin{equation}
		q_R = q(x=\tfrac{R}{a}) \equiv \delta N. 
	\end{equation}
	Consider further the eigenvalues $\xi$ of the differential operator $\hat{T}$
\begin{equation}
	\hat{T}(x) q(x) = \xi q(x).
\end{equation}
Equation (\ref{eq:eig_lx}) gives directly the relation between the eigenvalues $\lambda$ of the Kernel and the eigenvalues $\xi$ of the differential operator $\hat{T}$
	\begin{equation}\label{eq:lambda}
	\lambda_j = \xi_j + \frac{1}{R}\frac{q_R^2}{\int_0^{R/a} q(x)^2 dx},
\end{equation}
$j = 0,1,2,\ldots$.

	The eigenvalues $\lambda$ and the eigenfunctions $q$ can be calculated numerically. We first calculate the eigenvalues $\xi_j$ and eigenfunctions $q_j$ of the differential operator $\hat{T}$
	and then the eigenvalues $\lambda_j$ from
	(\ref{eq:lambda}).

	The radius $R_{\rm c}$ at which the first eigenvalue becomes zero $\lambda_0 = 0$ designates the minimum radius where the grand-canonical gravitational instability sets in, i.e. the critical turning point of stability, from stable to unstable configurations. Any mode $j$ with 
	\begin{equation}
		\text{unstable modes:}\quad \lambda_j > 0 
	\end{equation}
	is unstable. At each equilibrium where a new eigenvalue becomes zero, a new unstable mode is added, and in this sense this equilibrium is also a turning point. As is evident from (\ref{eq:lambda}), a mass increase $\delta N = q_R >0$ destabilizes the system. This was intuitively expected. What may be counter-intuitive is that the system gets destabilized also by a loss of mass $\delta N <0$. This, nevertheless, is consistent with mass conservation globally. 
	
	In order to solve the system numerically we perform the change of variables
	\begin{align}\label{eq:2_rescale}
		&w = \frac{n_{\rm b}}{n_0}\frac{1}{(\delta n_0/n_0)}\frac{\delta n}{n},\quad
		{\rm v} = \left(\frac{n_{\rm b}}{n_0}\right)^{1/2} \frac{1}{(\delta n_0/n_0)}
		\frac{\delta \nu}{6N_{\rm J}},
		\nonumber\\
		&\eta = \left(\frac{n_0}{n_{\rm b}}\right)^{1/2} \frac{a}{R_{\rm J}} x,\quad
		\sigma = \frac{R_{\rm J}^2}{a}\frac{n_{\rm b}}{n_0} \xi
	\end{align}
	and solve the system
	\begin{eqnarray}\label{eq:uQ_eq}
		\frac{d w(\eta) }{d\eta} &=& \frac{4\pi^2}{3}\left(\sigma - \frac{1}{\eta^2} \right) {\rm v}(\eta)
		\\
		\frac{d {\rm v}(\eta) }{d\eta} &=& 3 \eta^2 e^{-y(\eta)} w(\eta)
	\end{eqnarray}
	with boundary conditions
	\begin{equation}\label{eq:uQ_bound}
		w(0) = \frac{n_{\rm b}}{n_0},\quad
		w(\eta_R) = 0,\quad
		{\rm v}(0) = 0,
	\end{equation}
	where $\delta n_0 \equiv \delta n (0)$, $\eta_R \equiv \eta(x=R/a)$. For perturbations of the homogeneous equilibria we have $n_{\rm b}/n_0 = 1$, $y=0$.
	
	We stress that the perturbations are valid for any $\delta n_0/n_0$, arbitrarily small as is evident from the scalings (\ref{eq:2_rescale}). 
	Let us now calculate the eigenvalues $\sigma_j$, and through them the eigenvalues $\lambda_j$, and the corresponding eigenfunctions $(\delta n/n)_j$, for homogeneous and inhomogeneous equilibria.

		\subsubsection{Instability of homogeneous equilibria}
	
	The thermodynamic stability is governed by the eigenvalue problem (\ref{eq:K_eig}), equivalently expressed by the non-dimensional system (\ref{eq:uQ_eq}), (\ref{eq:uQ_bound}). We consider here `grand canonical' perturbations (allowing for a total mass variation) about the homogeneous equilibrium states and in the next subsection about the inhomogeneous ones.

	In Table \ref{tab:hom_eig} are depicted the eigenvalues of perturbations for those homogeneous states (at several radii $R$ of the perturbation) that a new unstable mode is added, for the first three modes.
	The critical turning point of stability occurs at
	\begin{equation}
		R_{\rm c} = 0.274 R_{\rm J},
	\end{equation}
	corresponding to critical mass
	\begin{equation}
		M_{\rm c} = 0.124 {\rm M}_{\rm J}.
	\end{equation}
	This is the minimum radius for which the zeroth perturbation mode $\lambda_0$. with zero nodes of the eigenfunction, becomes unstable. The eigenfunction $\delta n/n$ of this mode is depicted in \ref{fig:dn_R1_hom}. 
	
	From $R_{\rm c}$ until the radius $R_1 = 0.621 R_{\rm J}$ only the zeroth mode is unstable, whereas at $R_1$ there is a second turning point, where the first mode $\lambda_1$ becomes unstable. In Figure \ref{fig:dn_R2_hom} we depict the eigenfunctions of the two unstable modes. We find that the over-density within the first node contains $\sim 20\% $ of the total mass. 
	
	From $R_{\rm 1}$ until the radius $R_2 = 0.956 R_{\rm J}$ the first two modes are unstable, whereas at $R_2$ there occurs a third turning point, where the second mode $\lambda_2$ becomes unstable. In Figure \ref{fig:dn_R3_hom} we depict the eigenfunctions of the three unstable modes at this third turning point. 
	Remarkably, the zeroth mode at $R_2$, and any $R \geq R_2$ we checked, has one node (instead of noone as for smaller radii). 
	We find that the over-density within the zeroth node contains $\sim 30\% $ of the total. This suggests that for $R \geq R_{\rm J}$ thermodynamic instability implies a structure with a central object hosting $30\%$ or less of total mass. 
	
		Considering an intuitive description of the grand canonical gravitational instability, we remark that the system not only is in contact with the heat bath, dissipating away any excess of heat that would contribute to a balancing thermal pressure, but also being an open system, any pressure gradient --equivalently density gradient-- that could balance gravity, attracts more mass from the outer regions eliminating its balancing scope. This is a grand-canonical gravitational instability of both thermal and dynamical nature.
	The effective inertial force due to expansion, i.e. the apparent force in the comoving frame, 
	while crucial for allowing the homogeneous equilibria to exist, is insufficient to hold the collapse since it is more significant for the outer parts of the volume, 
	as it grows with $r$. On the contrary, above a certain lengthscale ($\sim R_{\rm J})$, its role is destabilizing as the instability may proceed in a core-halo type of structure where the outward parts become separated from the central core (as in the unstable eigenfunction of $\lambda_2$ in Figure \ref{fig:dn_R3_hom}).
	
	Above a certain lengthscale, a pressure gradient is required to balance the gravitational attraction, which gives rise to inhomogeneous equilibria. Nevertheless, they are also unstable, as we will immediately calculate in detail.
	
	\subsubsection{Instability of inhomogeneous equilibria}	
	
			\begin{figure}[tb]
		\centering
		\includegraphics[width=0.9\columnwidth]{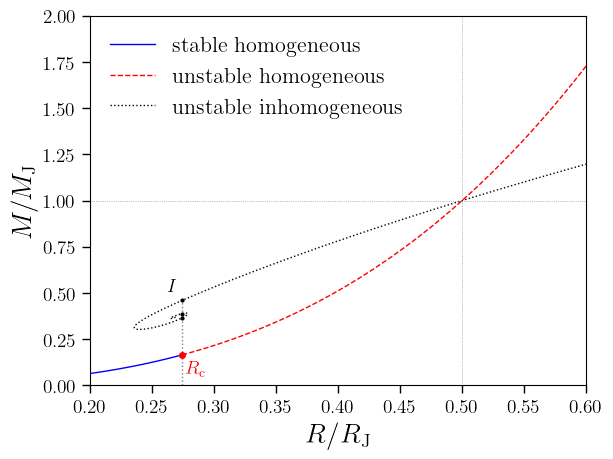}  
		\caption{Series of thermodynamic equilibria, expressed by total mass $M(R)$ contained within radius $R$, of self-gravitating monoatomic gas in an expanding background in the grand canonical ensemble.}
		\label{fig:MR_series}
	\end{figure}
	
	\begin{figure}[tb]
		\centering  
		\begin{subfigure}{\columnwidth}
			\centering  
			\includegraphics[width=0.85\textwidth]{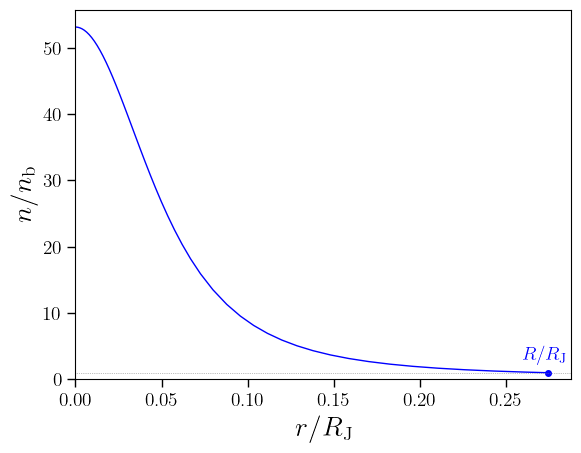} 
			\captionsetup{justification=centering} 
			\caption{Density profile.}
			\label{fig:n_r_I1}
		\end{subfigure}
		\begin{subfigure}{\columnwidth}
			\centering 
			\includegraphics[width=0.85\textwidth]{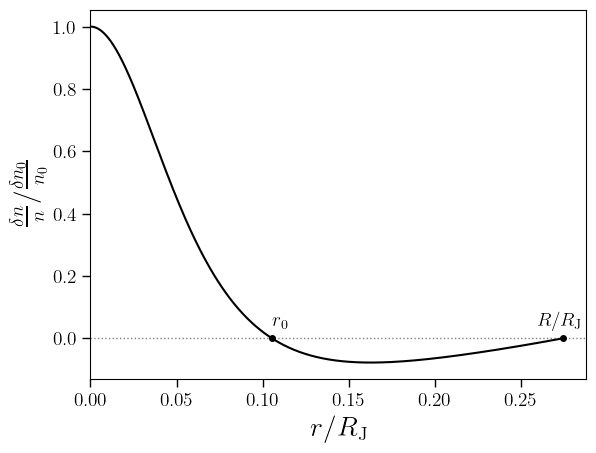} 
			\captionsetup{justification=centering}
			\caption{Unstable mode.}
			\label{fig:dn_R_I1}
		\end{subfigure}
		\caption{Panel (a): The density profile of the inhomogeneous equilibrium $I$, which is defined in Figure \ref{fig:MR_series};
			Panel (b): The eigenfunction of the first unstable mode of $I$ corresponding to eigenvalue $\lambda_0 = 47.02$.}
		\label{fig:I1}
	\end{figure}
	
	The inhomogeneous equilibria\footnote{Interestingly, similar solutions are found in the completely different context of radiative cooling spheres without expanding background \cite{2004ApJ...607..879M}.} are the non-trivial solutions of Poisson equation (\ref{eq:poisson_comoving_n}). The series of inhomogeneous equilibria (total mass of solutions at several radii) correspond to the spiral depicted in Figure \ref{fig:MR_series}.
	For $R > 0.5 R_{\rm J}$ the density of any equilibrium is increasing outwards, that is $n_0/n_{\rm b} < 1$, for each equilibrium. 
	On the other hand the ratio $n_0/n_{\rm b}$ of equilibria grows indefinitely for $R < 0.5 R_{\rm J}$ as we move along the spiral towards its focal point. At the equilibrium of minimum radius $R_{\rm min} = 0.2356 R_{\rm J}$ it is $n_0/n_{\rm b} = 574$.
	We stress that no inhomogeneous isothermal spheres exist for $R<0.2356 R_{\rm J}$ in an expanding Universe.
	
	We have solved the eigenvalue problem (\ref{eq:K_eig}) for several radii of inhomogeneous equilibria, corresponding to a wide range $-9 < \ln(n_0/n_{\rm b}) < 9$.
	We were always able to identify at least one instability mode at each equilibrium. This evidence along with the numerical behaviour of the eigenvalues we calculated --namely that more unstable modes are added in both cases we either decrease or increase $R$ departing from $R = 0.5 R_{\rm J}$-- leads us to conclude with reasonable certainty that the whole series of inhomogeneous equilibria are unstable under grand canonical perturbations. 
	
	In Figure \ref{fig:n_r_I1} we depict the density distribution of the less unstable inhomogeneous equilibrium corresponding to the critical radius $R_{\rm c}$ designating the onset of the instability of homogeneous states. This inhomogeneous equilibrium is marked as point $I$ in Figure \ref{fig:MR_series}.
	Its total mass is
	$M_{\rm I} = 233,715 {\rm M}_\odot$ and the density ratio is $n_0/n_{\rm b} = 53$.
	In Figure (\ref{fig:dn_R_I1}) we show the eigenfunction of the zeroth mode $\lambda_{{\rm I},0} = 47.02$, that is the first unstable mode.
	It contains one node. Every unstable mode in all inhomogeneous equilibria we considered contains at least one node. The mass within the over-density inside the first node of equilibrium $I$ is
	$M_{\rm I}(r<r_0) = 73,596 {\rm M}_\odot$, that is $30\%$ of the total mass. We encounter again the same ratio of central to outer mass as in the case of $R_2 \simeq R_{\rm J}$ homogeneous equilibrium. The instability of every inhomogeneous equilibrium suggests a resulting structure of a central collapsed object hosting a significant ratio of the total mass.
	
		The instability of the inhomogeneous equilibria, we think, is in principle not different in nature than that of the homogeneous ones. The isothermal and fixed boundary pressure conditions are crucial. The thermal pressure (and its gradient in this case) is fixed, but gravitational attraction grows with mass.
	
	Since all inhomogeneous equilibria are almost certainly unstable under the perturbations considered, we shall not discuss them further. We only remark that their existence and their respective instabilities may dictate the direction towards which the instability of the homogeneous equilibria at the same radius shall evolve. However, this is speculative at this level of analysis and requires further investigation.
	
		\subsection{Canonical perturbations}\label{sec:stability_can}
	
	In section (\ref{sec:stability}) we calculated the condition (\ref{eq:2nd_order_inst}) for an instability to set in allowing for total mass variations $\delta N \neq 0$, corresponding to the grand canonical ensemble. Let us inspect here the case of `canonical' perturbations with $\delta N = 0$, which correspond to a stability analysis in the canonical ensemble. Recall that ensembles are not equivalent in the statistical mechanics of self-gravitating systems, regarding their stability properties \citep{1990PhR...188..285P}. This steams from the non-additive character of gravity and is explicitly evident in the boundary conditions of the respective perturbations. Particularly, in our case the term $\delta N$ in equations (\ref{eq:2nd_order}) and (\ref{eq:2nd_order_inst}) should be zero in the canonical ensemble. This renders equation (\ref{eq:2nd_order}) a variation of the Helmholtz free energy and transforms (\ref{eq:2nd_order_inst}) to
	\begin{align}\label{eq:d2F}
		&- \int_0^{R/a} \delta \nu(x) \times
		\nonumber \\
		&\left\lbrace
		\frac{Gm^2 \beta}{a^2 x^2}
		+
		\frac{d}{dx}\left(\frac{1}{4\pi a^4 x^2 n(x)}\frac{d}{dx} \right)
		\right\rbrace
		\delta \nu(x)
		a dx
		< 0 .
	\end{align} 
	This is the condition for an instability to set in, in the canonical ensemble. A turning point of stability occurs when the above inequality becomes equality
	\begin{equation}\label{eq:can_cond}
		\frac{Gm^2 \beta}{x^2}\delta \nu
		+
		\frac{d}{dx}\left(\frac{1}{4\pi a^2 x^2 n(x)}\frac{d}{dx} \right)\delta\nu = 0.
	\end{equation}
	Note that this condition gives the condition for the onset of dynamical Jeans instability for homogeneous equilibria in the limit of slow expansion, as we show  in \ref{app:inst_condition} (equation (\ref{eq:app:dyn_eig})). Therefore, thermodynamic gravitational instability in the canonical ensemble for homogeneous equilibria in expanding background implies (Jeans) dynamical instability. This is known to be true also in Newtonian gravity without an expanding background \cite{2013A&A...552A..37S} and for certain conditions in General Relativistic systems \cite{2013CQGra..30k5018R,2015CQGra..32k9501R,Green_2014}. This is not the case in the grand canonical ensemble, as discussed further in \ref{app:inst_condition}.

	For homogeneous equilibria $n = n_{\rm b}$ it gives
	\begin{equation}\label{eq:d2F_can}
		\frac{1}{\tilde{x}^2} q
		+
		\frac{1}{4\pi^2}
		\frac{d}{d\tilde{x}}
		\left(\frac{1}{\tilde{x}^2 }\frac{d}{d\tilde{x}}\right) 
		q
		= 0 ,
	\end{equation} 
	with initial conditions
	\begin{equation}\label{eq:ic_can}
		q(0) = 0,\quad 
		q^\prime(0) = 0,
		w(0) = w_0, 
	\end{equation}
	and boundary condition
	\begin{equation}\label{eq:bc_can}
		w(R_{\rm cc}) = 0,
	\end{equation}
	where $R_{\rm cc}$ denotes the critical radius in the canonical ensemble and we define
	\begin{equation}
		w \equiv \frac{\delta n}{n_{\rm b}},\quad
		\tilde{x} = \frac{x}{R_{\rm J}/a}.		
	\end{equation}
	The transformations
	\begin{equation}
		s = 2\pi \tilde{x},\quad
		{\rm y} = \tilde{x}^{-\frac{3}{2}} q
	\end{equation}
	bring equation (\ref{eq:d2F_can}) to the Bessel form
	\begin{equation}
		s^2 {\rm y}(s)^{\prime\prime} + s {\rm y}(s)^{\prime} + \left(s^2 - (\tfrac{3}{2})^2\right) {\rm y}(s) = 0.
	\end{equation}
	The solution is given with respect to the Bessel functions of the first and second kind
	\begin{equation}
		q(s) = s^{\frac{3}{2}} \left(C_1 J_{\frac{3}{2}}(s) + C_2 Y_{\frac{3}{2}}(s) \right),
	\end{equation}
	where 
	\begin{align}
		&J_{\frac{3}{2}}(s)  = \sqrt{\frac{2}{\pi s}}\left(\frac{\sin(s)}{s} - \cos(s) \right)
		\\
		&Y_{\frac{3}{2}}(s)  = -\sqrt{\frac{2}{\pi s}}\left(\frac{\cos(s)}{s} + \cos(s) \right) .
	\end{align}
	It is now straightforward to calculate $C_1$, $C_2$ using the initial conditions (\ref{eq:bc_can}) and get finally
	\begin{align}
		&\frac{\delta \nu(\tilde{x})}{\delta n(0)/n_{\rm b}} = n_{\rm b} \frac{4}{3}\pi R_{\rm J}^3\frac{3}{(2\pi)^3} \left(\sin(2\pi\tilde{x}) - 2\pi \tilde{x}\cos(2\pi\tilde{x}) \right)
		\\
		& \frac{\delta n(\tilde{x})/n_{\rm b}}{\delta n(0)/n_{\rm b}} = \frac{\sin(2\pi\tilde{x})}{2\pi\tilde{x}} .
	\end{align}
	The boundary condition (\ref{eq:bc_can}) specifies the critical radius $R_{\rm cc}$ and the secondary turning points of stability
	\begin{equation}
		R_{{\rm can},i} = \frac{i}{2} R_{\rm J}, \; i=1,2,\ldots .
	\end{equation}
	The critical point in the canonical ensemble is therefore
	\begin{equation}
		R_{{\rm cc}} = \frac{1}{2} R_{\rm J},
	\end{equation}	
	with
	\begin{equation}
		M_{\rm cc} = M_{\rm J} .
	\end{equation}
	At the critical point $R_{\rm cc}$ the homogeneous series and the inhomogeneous series cross, as in Figure \ref{fig:MR_series}. We conclude
	that, under canonical perturbations, for $R<0.5R_{\rm J}$ the homogeneous series is stable and the inhomogeneous unstable. At the critical point a change of stability occurs for both series. For $R > 0.5 R_{\rm J}$ the inhomogeneous series is stable and the homogeneous series unstable.
	
	\section{Application in astrophysics}\label{sec:astrophysics}
	
	\begin{table*}[tb]
		\begin{center}
			\begin{tabular}{c c | c  c  c c }			
				$R(z=1100)$
				&				
				$R(z=300)$
				&				
				$M_{\rm tot}$
				&				
				$M(r<r_0)$
				&				
				$M(r<r_1) $
				&				
				$M(r<r_2) $
				\\
				\toprule
				$15 {\rm pc}$
				&				
				$60 {\rm pc}$
				&				
				$83,883 {\rm M}_\odot$
				&				
				$83,883 {\rm M}_\odot$
				&				
				\color{green}{stable}
				&
				\color{green}{stable}
				\\
				\midrule
				$33 {\rm pc}$
				&				
				$122 {\rm pc}$
				&				
				$974,968 {\rm M}_\odot$
				&				
				$974,968 {\rm M}_\odot$
				&				
				$197,021 {\rm M}_\odot$
				&
				\color{green}{stable}
				\\
				\midrule
				$51 {\rm pc}$
				&				
				$188 {\rm pc}$
				&				
				$3,548,025 {\rm M}_\odot$
				&
				$1,057,648 {\rm M}_\odot$
				&				
				$480,753 {\rm M}_\odot$
				&
				$280,336 {\rm M}_\odot$
			\end{tabular}
			\caption{The first three critical turning point radii at the two limiting redshift values, along with the total mass and the mass within radii $r_0$, $r_1$, $r_2$, for grand canonical perturbations about the homogeneous state. Each one of these radii denotes the radius of the first node of the respective eigenfunction, as in Figure \ref{fig:dn_R_hom}.}
			\label{tab:hom_astr}
		\end{center}
	\end{table*}
	
	The, so-called, dark ages begin after the recombination era has ended, $\sim 400,000$ years after the big bang, at redshift $z \sim 1100$, according to the standard model of Cosmology \citep{2020A&A...641A...6P,1993ppc..book.....P,1999ApJ...523L...1S,2000ApJS..128..407S,2001PhR...349..125B}.
	Baryonic matter is in the form primarily of neutral atomic hydrogen and its temperature is coupled with the Cosmic Microwave Background (CMB) radiation until cosmic time $\sim 10{\rm Myr}$, that is $z \sim 137$ \citep{1999ApJ...523L...1S}.
	From then on, baryons cool at a faster rate than radiation due to Universe's expansion. Matter remains nearly homogeneous to all scales up to cosmic time $\sim 100{\rm Myr}$, at redshift $z \sim 30$, as non-linear gravitational collapse of dark matter was triggered in overdense regions and the first small-scale structures, mini-halos and protogalaxies, begin to form. 
	
	Therefore, within the window $1100 \gtrsim z \gtrsim 137$ the dominant baryonic component, namely atomic Hydrogen, is spatially isothermal with a temperature that cools at expansion timescales with a rate equal to that of CMB radiation $T(z) \propto 1+z$. 
	The thermal relaxation timescale is negligible (see \ref{app:t_rel}) with respect to the expansion timescale, which is of the order of the dynamical timescale, therefore the temperature $T(z)$, cosmic scale factor $a(z)$, matter density $\rho(z)$ are adiabatic invariants, that is they can be considered constant at thermal relaxation timescales. 
	The density is homogeneous at least down to $\delta \rho/\rho \sim 10^{-5}$ \citep{1993ppc..book.....P}. 
	Any density perturbation within a restricted volume should equal the background homogeneous density value.
	These conditions are consistent with our assumptions (see also \ref{app:hydrostatic}) that lead to the thermodynamic analysis we developed previously.

	A crucial question is whether the instability of the homogeneous equilibria has the time to evolve from $z=1100$ down to $z=137$, at the time when the gas decouples thermally from the CMB (Compton scattering becomes ineffective \citep{2001PhR...349..125B}). Assuming the age of the Universe is
	\begin{equation}\label{eq:t-z}
		t(z)=\frac{1}{H_0}\int_0^{1/(1+z)}\frac{dx}{x\sqrt{\Omega_\Lambda+\Omega_kx^{-2}+\Omega_mx^{-3}+\Omega_\gamma x^{-4}}},
	\end{equation}
	with $H_0=67.4{\rm km}/{\rm s}/{\rm Mpc}$, $\Omega_{\rm m}=0.315$, $\Omega_\Lambda = 0.685$, $\Omega_\gamma = 5\cdot 10^{-5}$, $\Omega_k = 0$ \citep{2020A&A...641A...6P}, the available time window for the instability to evolve is 
	$t(137) - t(1100) \simeq 10 {\rm Myr}$.
	
	It is reasonably expected \cite{Spitzer1978} that the instability evolves over a time of the order of the (free fall) dynamical timescale $t_{\rm dyn} = \sqrt{3/32\pi Gm\,n_{\rm b}}$ (this is also suggested by the dynamic equation (\ref{eq:app:eul})).
	Assuming $n_{\rm b} = n_{\rm H} \simeq 1.9\cdot 10^{-7}{\rm cm}^{-3} (1+z)^3$, the dynamical timescale  gives
	\begin{equation}\label{eq:t_dyn-z}
		t_{\rm dyn}(z) = 1.2{\rm Myr} \left(\frac{1+z}{1000}\right)^{-3/2}.
	\end{equation}
	Since the instability proceeds over the dynamical timescale which dependds on the redshift as above, and since the gas is isothermal only until redshift $z\simeq 137$, we are looking for the minimum redshift $z_{\rm dyn}$ which corresponds to a time interval $\Delta t$ with respect to $z=137$, that is equal to one dynamical timescale, that is $\Delta t \equiv t(137) - t(z_{\rm m}) = t_{\rm dyn}(z_{\rm m})$.
	From equations (\ref{eq:t-z}) and (\ref{eq:t_dyn-z}) we can infer this minimum redshift $z_{\rm m}$, 
	for which the instability has sufficient time to proceed isothermally and get
	$z_{\rm m} = 299$.
	Thus, the instability may develop only if initiated within the window
	\begin{equation}
		300 \lesssim z \lesssim 1100 .
	\end{equation}
		Note that during this period it is $M_{\rm J} = {\rm const.}$, i,e, the Jeans mass is independent from redshift, because the dependences $n_{\rm b} \propto (1+z)^3$ and $T\propto (1+z)$ cancel out.
	The temperature within this cosmic time window is $T = 1000-3000{\rm K}$. Therefore, the three radii of the critical turning point and the first two secondary turning points of Table \ref{tab:hom_eig} are bounded as follows, 
	$15{\rm pc} \leq R_{\rm c} \leq 60 {\rm pc}$ for $M_{\rm c} = 0.84\cdot 10^5{\rm M}_\odot$, 
	$33{\rm pc} \leq R_{\rm 1} \leq 122 {\rm pc}$ for $M_{\rm 1} = 9.75\cdot 10^5{\rm M}_\odot$,
	$51{\rm pc} \leq R_{\rm 2} \leq 188 {\rm pc}$ for $M_{\rm 2} = 3.55\cdot 10^6{\rm M}_\odot$.
	The critical radius as a function of redshift is
	\begin{equation}
		R_{\rm c} = 16.22{\rm pc} \left(\frac{1+z}{1000} \right)^{-1}
	\end{equation}

	In Table \ref{tab:hom_astr} we show the first three turning points of stability along with the mass contained within the first node for each mode.
	Perturbations with lengthscales $R > 0.96 R_{\rm Jeans}$ at the aforementioned redshift window suggest a centrally condensed system with a core comprising $\sim 30\% $ of the total mass of the cloud. Such structures, with extreme mass over-ratio, are nevertheless consistent with tantalizing recent observations of AGNs at high redshift, which contrary to local relations --that indicate MBH to galactic stellar mass ratios ${\rm M}_{\rm MBH} / {\rm M}_\star < 0.3\%$ or even lower-- show unexpectedly, extreme high ratios ${\rm M}_{\rm MBH} / {\rm M}_\star \sim 30\%$ \citep{2024NatAs...8..126B,2024arXiv240610329D,2024arXiv240303872J,2024ApJ...965L..21K,2023ApJ...957L...7K} in contrast with the standard hierarchical merging paradigm.
	
	The other key issue is the evolution of the destabilized gaseous cloud. This requires different types of extensive analyses, but let us discuss some basic points here.
	During this cosmic era perturbations grow only linearly with time. At $z\lesssim 137$, when the gas decouples thermally from the CMB radiation and cools at higher rates, the cloud can not have grown more than $\delta n/n \sim 1$, while radiation pressure may have even prevented such high values. We emphasize that while we assumed a homogeneous dark matter component, in order to focus on the gas component and inspect the scenario least susceptive to gravitational instability, it seems natural that the perturbation of the gas will develop within a dark matter over-density. It does not seem probable that non-linear small-scale structure formation gets ignited before dark matter becomes non-linear gravitationally unstable (estimated at $z\sim 30$ \citep{2001PhR...349..125B}) though re-examining established conclusions may not be unjustified in the light of new observational data that are in tension with such conclusions. Nevertheless, it is possible that due to the thermodynamic gravitational instability the non-linear small scale structure formation is ignited a little earlier and/or develops faster than estimated in the standard model.
	
	Finally, let us comment on potential fragmentation of the destabilized clouds. The primary molecular cooling mechanisms, that could drive fragmentation if the cooling timescale $t_{\rm cool}$ becomes about equal or less than the dynamical timescale 
	$t_{\rm cool} \lesssim t_{\rm dyn}$ \citep{1977MNRAS.179..541R,1985ApJ...298...18F}, are not only suppressed at these redshifts, but also are exactly these mechanisms that couple the gas with the CMB radiation. Therefore they are not in effect ``cooling'' mechanisms but regularizing ones. These are in particular Hydrogen line cooling, cooling by molecular Hydrogen and Compton cooling \citep{1997ApJ...474....1T}.
	For example, in the Compton case $d T/ dt|_{\rm Compton} \propto T_\gamma - T_{\rm gas}$ (see \citep{1997ApJ...474....1T} for the rest mechanisms). Wherever $T_{\rm gas} < T_{\gamma}$ the gas is heated and if $T_{\rm gas} > T_{\gamma}$ the gas is cooled, maintaining the CMB radiation temperature. 
	Also, the formation of molecular Hydrogen via $H$, ${\rm e}^{-}$ interactions, that radiates energy, is suppressed at $z \sim 1100$ because of the low electron density, with a fraction $\sim 10^{-4}$ of $H$ \citep{2000ApJS..128..407S}. Still, the generated $\gamma$ and molecular Hydrogen would only serve to couple the gas' temperature with that of CMB radiation.

	\section{Conclusion}\label{sec:conclusion}
	
	We developed the thermodynamics of self-gravitating isothermal gas in an expanding homogeneous background.
	The apparent inertial force at comoving coordinates allows not only for homogeneous equilibria to exist, but also inhomogeneous ones above a radius $R > 0.236R_{\rm J}$. 
	For $R >0.5 R_{\rm J}$ these have outwards increasing density. 
	Qualitatively, at larger lengthscales, and correspondingly greater masses, the homogeneous configuration is not sufficient to hold the collapse because the pressure is constant throughout the volume, but gravitation grows with mass. A pressure gradient is required to balance gravity giving rise to inhomogeneous configurations at large lengthscales.
	
	Our theoretical analysis has revealed an, as yet unnoticed, thermodynamic gravitational instability for both the homogeneous and inhomogeneous series of equilibria.
	For grand canonical perturbations at fixed boundary density, equal to the homogeneous background, 
	the whole inhomogeneous series is unstable.
	Therefore, no inhomogeneous isothermal spheres exist with radius $R<0.236 R_{\rm J}$ if the Universe expansion is taken into account and those that exist above this radius are unstable for perturbations that allow the exchange of mass with the environment. 
	
	Most remarkably, an instability sets in also at the homogeneous series at radius $0.274 R_{\rm J}$ for these grand canonical perturbations. At lengthscales greater than this critical value the homogeneous gas is unstable. 
	
	For perturbations that preserve the total mass (and temperature), corresponding to the canonical ensemble, the homogeneous series is stable and the inhomogeneous unstable for $R<0.5 R_{\rm J}$ and the reverse for $R > 0.5 R_{\rm J}$. 
	
	A potential astrophysical application refers to the atomic Hydrogen gas during the dark ages time window $1100 \gtrsim z \gtrsim 300$, when the Universe is spatially homogeneous at all scales, and the baryonic component is coupled thermally to the CMB radiation, allowing for isothermal perturbations to develop.
	We predict that sub-galactic-scale gaseous clouds with masses $M \geq 83,883 {\rm M}_\odot$
	are susceptible to the grand-canonical gravitational instability with critical lengthscales 
	$R_{\rm c}(z=110) = 15 {\rm pc}$, $R_{\rm c}(z=300) = 60 {\rm pc}$.
	At lengthscales 
	$R \gtrsim R_{\rm J}$, 
	corresponding to 
	$\sim 50-190 {\rm pc}$
	 in the referring cosmic time window, the instability mode of the destabilized clouds suggests the development of a central core hosting $\sim 30\%$ of the total mass, in analogy with recent intriguing observations of AGN at high redshift with extreme over-ratios
	 $M_{\rm MBH}/M_\star$ 
	 \citep{2024NatAs...8..126B,2024arXiv240610329D,2024arXiv240303872J,2024ApJ...965L..21K,2023ApJ...957L...7K}.
	
	Future developments could include in the theoretical side a generalization of the thermodynamic framework to the general relativistic regime that may apply to even earlier stages of Universe evolution, and in the astrophysical side the calculation of the evolution of the identified clouds at lower redshifts. 
	
	\section*{Acknowledgements}
	\noindent	
	ZR is supported by the European Union's Horizon Europe Research and Innovation Programme under the Marie Sk\l{}odowska-Curie grant agreement No.~101149270--ProtoBH.
	
	\bibliography{iso_inst}
	\bibliographystyle{apsrev4-2}	
	
	\appendix

\section{Thermal relaxation timescale}\label{app:t_rel}

	The thermal relaxation timescale of atomic Hydrogen is 
\begin{equation}
	t_{\rm rel} = \sqrt{\frac{\pi m}{8T}}\frac{1}{n_{\rm b}\sigma_H}
\end{equation}
The cross section $\sigma_H$ shall depend on temperature at high temperatures \citep{1954mtgl.book.....H}, where the effective interaction lengthscale decreases because atoms can overcome potential energy barriers more readily. Simulations suggest a dependence for atomic Hydrogen --the so-called variable soft spheres model \citep{1994mgdd.book.....B}-- of the form
$
\sigma_{H} \simeq \sigma_0 \left(T_0/T\right)^{2/3}
$ 
(see also \citep{1954mtgl.book.....H,1961PSS....3..236D,1970mtnu.book.....C}).
Using as reference values the Lennard-Jones cross section $\sigma_{\rm LJ} = 1.78\cdot 10^{-19}m^2$ at $T = 273{\rm K}$ \citep{1954mtgl.book.....H} we get for the relaxation timescale
\begin{equation}
	t_{\rm rel}(z) = 5.7\cdot 10^{-7}{\rm Myr} \left(\frac{1+z}{1000} \right)^{-\frac{17}{6}}.
\end{equation}
Therefore, considering the dynamical timescale (\ref{eq:t_dyn-z}) we have in the window $1100 \lesssim z \lesssim 300$ that $t_{\rm rel}/t_{\rm dyn} \lesssim 10^{-6}$.
	
	\section{Hydrostatic formulation}\label{app:hydrostatic}
	
	We remark that in the calculation of thermodynamic equilibria in section \ref{sec:equilibria} we did not use the ideal gas equation of state $P(r) = n(r)/\beta$, where $P$ is the pressure. We did not even refer to pressure, but still we got the ideal gas mean kinetic energy $K = 3N/2\beta$ as in eq. (\ref{eq:K}). We also got the Boltzmann factor in the phase space distribution without any reference to pressure. Apparently, the ideal gas equation of state is implicitly incorporated in the analysis. This is a result of Boltzmann entropy (\ref{eq:S_def}) that encapsulates the mean field approximation. This expression presumes that short distance effects are negligible, naturally leading to an ideal gas equation of state \citep{2003FoPh...33..223K}.
	
	Furthermore, let us show that Poisson equation (\ref{eq:poisson_comoving_n}) with the Boltzmann factor (\ref{eq:n_bol}) are consistent with the expanding background at the cosmic era considered here.
	We clarify at this point that `hydrostatic', just like `thermodynamic equilibrium', in our context refers to configurations static within thermal relaxation timescales, which are negligibly smaller w.r.t. the cosmic expansion timescale. Now, let us first verify that all fluid equations, including mass conservation, are consistent.
	
	The continuity and Euler equations, become in comoving coordinates for an ideal gas, following Peebles (see equations (9.15), (9.17a) in \citep{1980lssu.book.....P}),
	\begin{align}
		\label{eq:app:fluid_cont}
		&\frac{\partial \rho}{\partial t} + 3\frac{\dot{a}}{a}\rho+\frac{1}{a}\nabla(\rho \bm{v}) = 0
		\\
		\label{eq:app:eul}
		&\frac{\partial \bm{v}}{\partial t} + \frac{1}{a}(\bm{v}\cdot \nabla)\bm{v}+ \frac{\dot{a}}{a}\bm{v} = -\frac{1}{\rho a}\nabla P - \frac{1}{a}\nabla \phi
	\end{align}
	where $\bm{v}$ is the velocity relative to the backgound defined in (\ref{eq:v}), $\phi$ satisfies the Poisson equation (\ref{eq:poisson_comoving_n}) and $\rho \equiv mn$. The mass conservation in the hydrostatic case $\bm{v}=0$ gives 
	\begin{equation} 
		\rho \propto a^{-3} 
	\end{equation}
	as it should be. The Poisson equation consequently implies 
	\begin{equation}
		\phi\propto a^{-1}.
	\end{equation}
	Now, in the hydrostatic case $\bm{v}=0$, equation (\ref{eq:app:eul}) reads
	\begin{equation}\label{eq:hydrostatic}
		\nabla P = -\rho \nabla \phi
	\end{equation}
	which implies
	\begin{equation}
		P \propto a^{-4}.
	\end{equation}
	Therefore, for the isothermal, ideal gas equation of state 
	\begin{equation}\label{eq:ideal_eos}
		P = \frac{\rho}{\beta}
	\end{equation}
	we get
	\begin{equation}
		\beta \propto a.
	\end{equation}
	Since $a\propto (1+z)^{-1}$ this is exactly the correct scaling for as long as the gas is thermally coupled with the CMB radiation for which $T\propto (1+z)$.
	
	Furthermore, the solution of the hydrostatic equation (\ref{eq:hydrostatic}) for the isothermal, ideal gas equation of state (\ref{eq:ideal_eos}) gives the Boltzmann factor
	$\rho = \rho_{\rm b} e^{-m\beta (\phi - \phi_R)}$,
	and 
	\begin{equation} 
		\beta \phi \propto a\cdot a^{-1} 
	\end{equation}	
	is not evolving as expected for a self-consistent analysis.   
	Density evolves in cosmic times as $\rho\propto \rho_{\rm b} \propto a^{-3}$, and not exponentially w.r.t. $a$, despite the Boltzmann factor, even for the inhomogeneous equilibria.

	\section{Instability Condition}\label{app:inst_condition}

	We derive the grand canonical instability condition (\ref{eq:2nd_order_inst}). Recall the entropy (\ref{eq:S_def}) as a functional of the phase space density $f$
	\begin{equation}\label{eq:app:S_def}
		S[f] = - \iint f(\bm{x},\bm{p}) \ln f(\bm{x},\bm{p}) d\bm{\tau}.
	\end{equation}
	For a perturbation of the phase space density it is (for fixed volume)
	\begin{equation}
		\delta^{(2)} S = - \iint \left. \frac{\partial^2 }{\partial f^2}(f\ln f)\right|_{\rm eq}\,(\delta f)^2 d\bm{\tau}
	\end{equation}
	which gives 
	\begin{equation}
		\delta^{(2)}S = -\frac{1}{2} \iint \frac{(\delta f)^2}{f} d\bm{\tau},
	\end{equation}
	where the equilibrium $f$ is given by equation (\ref{eq:feq_n}).
	We consider only perturbations in the density $n$ and since $f$ is factorized as $f = {[\text{function of }p]}\times n(x)$ and considering the integral of $f$ in $d^3{\bm p}$ given in equation (\ref{eq:n_def}), we get
	\begin{equation}
		\delta^{(2)}S = -\frac{1}{2} a^3 \int \frac{(\delta n)^2}{n} d^3\bm{x},
	\end{equation}
	which is just equation (\ref{eq:dS2}). 
	
	Now, since the kinetic energy (\ref{eq:K}) is linear in $f$, its second order variation is zero. The $\delta^{(2)}E$ is determined by the variation in the potential energy (\ref{eq:U_def}), which by integrating out the momenta through (\ref{eq:n_def}) gives
	\begin{align}\label{eq:app:U_def}
		U = &-\frac{1}{2} Gm^2a^5\iint \frac{n(\bm{x}) n(\bm{x}^\prime) }{\left|\bm{x} - \bm{x}^\prime \right|} d\bm{x} d\bm{x^\prime} +
		\nonumber \\
		&+ a^3 m\int \phi_{\rm ext}(\bm{x}) n(\bm{x}) d\bm{x}.
	\end{align}
	Considering that $\phi = \phi_{\rm N} + \phi_{\rm ext}$ and the definition of the Newtonian potential (\ref{eq:phi_N_P}) we get
	\begin{equation}
		U = \frac{a^3}{2} \int m\left(\phi+ \phi_{\rm ext}\right) n d\bm{x}, 
	\end{equation}
	which gives directly
	\begin{equation}
		\delta^{(2)} U = \frac{a^3}{2} \iint m\,\delta \phi \,\delta n \, d\bm{x}.
	\end{equation}
	Note also that the total number of particles (total mass) is linear in $n$ therefore $\delta ^{(2)}N = 0$.
	
	Thus, the variation $\delta^{(2)}J \equiv	\beta \delta^{(2)} E  - \delta^{(2)} S - \alpha \delta^{(2)}  N$ is equal to 
	\begin{equation}\label{eq:app:d2J_1}
		\delta^{(2)}J = \frac{a^3}{2} \int \left( m\beta \,\delta\phi\, \delta n + \frac{(\delta n)^2}{n} \right) d^3{\bm x}.
	\end{equation}
	We have that
	\begin{align}
		&\delta n = \frac{1}{4\pi a^3 x^2} \frac{d(\delta \nu)}{dx},
		\\
		&\delta\nu(0) = 0,\quad
		\delta \nu(\tfrac{R}{a}) = \delta N,
		\\
		&\frac{d(\delta\phi)}{dx} = \frac{Gm}{ax^2}\delta\nu.
			\end{align}
	We integrate by parts equation (\ref{eq:app:d2J_1}), utilizing the above equations, to get consequtively
	\begin{align}\label{eq:app:d2J_2}
		\delta^{(2)}J &= \frac{a^3}{2} 4\pi \left\lbrace \int_0^{R/a} dx\,x^2\, m\beta \,\delta\phi\,  \frac{1}{4\pi a^3 x^2} \frac{d(\delta \nu)}{dx} 
		+ \right.\nonumber \\
		&\quad\left. + \int_0^{R/a} dx\, x^2\,\frac{\delta n}{n}\, \frac{1}{4\pi a^3 x^2} \frac{d(\delta \nu)}{dx}\right\rbrace
		\nonumber \\
		&= \frac{1}{2}\left\lbrace
		m\beta \delta\phi_R\, \delta N
		- \int_0^{(R/a)} dx\,\times
		\right.
		\nonumber \\
		&\quad 
		 \left.\left( \frac{Gm^2\beta}{ax^2}\delta\nu^2 + 
		 \delta\nu\frac{d}{dx}\left( \frac{1}{4\pi a^3 n \,x^2} \frac{d(\delta \nu)}{dx}\right)\right)
		 \right\rbrace ,
	\end{align}
	where $\phi_R \equiv \phi(\tfrac{R}{a})$ is the boundary potential given in (\ref{eq:phiR}),
	$\phi_R = -G\frac{mN}{R} + \phi_{\rm ext, R}$. Therefore for a fixed volume it is (the same result one may obtain by direct integration of $\delta\phi(\tfrac{R}{a})$ for $\phi$ given in (\ref{eq:phi_N_P}), (\ref{eq:phi_ext_P}))
	\begin{equation}
		\delta \phi_R = - G\frac{m}{R}\delta N.
	\end{equation}
	Substituting back into (\ref{eq:app:d2J_2}) we get
	\begin{align}\label{eq:app:d2J_3}
		\delta^{(2)}J = 
		&-\frac{1}{2}\left\lbrace \frac{Gm^2 \beta}{R}(\delta N)^2 
		+ \int_0^{R/a} dx\, \delta \nu(x) \times
		\right.
		\nonumber \\
		&\left. 
		\left(
		\frac{Gm^2 \beta}{a x^2} +
		\frac{d}{dx}\left(\frac{1}{4\pi a^3 x^2 n(x)}\frac{d}{dx} \right)
		\right)
		\delta \nu(x)
		\right\rbrace
	\end{align}
and the instability condition $\delta^{(2)}J < 0$ is just the condition (\ref{eq:2nd_order_inst}).

It is evident from equation ({\ref{eq:app:d2J_1}) that thermodynamic stability depends not only on the density variation $\delta n$, but also depends explicitly on the mass variation $\delta \nu$ through the potential term $\delta \phi$. On the contrary dynamical stability does not depend explicitly on mass. Particularly for our isothermal case with $\delta P = \delta n / \beta$, Peebles \cite{1980lssu.book.....P}, have shown by perturbing equations (\ref{eq:app:fluid_cont}), (\ref{eq:app:eul}) above, about homogeneous equilibria, that
the equation that determines the hydrodynamic evolution (see equations (10.2), (16.2) of \cite{1980lssu.book.....P})} is 
\begin{equation}\label{eq:app:dyn}
	4\pi G m \delta n + \frac{1}{a^2 m\beta n_{\rm b}}\nabla^2\delta n = \frac{\partial^2\delta n}{\partial t^2} + 2\frac{\dot{a}}{a}\frac{\partial \delta n}{\partial t}.
\end{equation}	
In the limit of slow expansion, and for perturbations
\begin{equation}
	\delta n(x,t) \sim e^{\sigma t}\delta n(x)
\end{equation}
this gives
\begin{equation}\label{eq:app:dyn_eig}
	4\pi G m \delta n + \frac{1}{a^2 m\beta n_{\rm b}}\nabla^2\delta n = \sigma^2 \delta n
\end{equation}	
with instability modes $\sigma^2 > 0$ and the onset of instability being triggered at $\sigma = 0$.

We can show that the onset thermodynamic condition (\ref{eq:can_cond}) in the canonical ensemble (that is derived from (\ref{eq:app:d2J_3}) above for $\delta N = 0$) implies the dynamical onset codition, by differentiating (\ref{eq:can_cond})
for $n=n_b$ and then substituting (\ref{eq:dn})
\begin{align}
	&Gm^2\beta \frac{d\delta\nu}{dx} + \frac{1}{4\pi a^2n_{\rm b}}\frac{d}{dx}\left\lbrace x^2 \frac{d}{dx} \left( \frac{1}{x^2}\frac{d\delta \nu}{dx}\right)\right\rbrace = 0
	\nonumber\\
	&\nonumber\\
	&4\pi Gm^2\beta a^2 n_{\rm b}\delta n + \frac{1}{x^2}\frac{d}{dx}\left( x^2 \frac{d}{dx} \delta n\right) = 0
	\nonumber\\
	\label{eq:app:Jeans}
	&4\pi Gm \delta n + \frac{1}{a^2 m \beta n_{\rm b}}\nabla^2 \delta n = 0.
\end{align}
This is Jeans criterion. A thermodynamic instability in the canonical ensemble in expanding background implies Jeans instability.

However, the grand canonical condition cannot be derived from (or lead to) the dynamical condition. Is is straightforward to verify that even if one substitutes in (\ref{eq:app:dyn_eig}) the equation (\ref{eq:dn}), i.e. $(\delta \nu)^\prime = 4\pi a^2 x^2 \delta n$, multiplies by $\delta \nu$ and integrates from $x$ to $R/a$ will not get the correct $\delta N^2$ term of (\ref{eq:app:d2J_3}). Apparently, the grand canonical and canonical conditions are not equivalent. This should be obvious from the start, because as we remarked the thermodynamic condition (\ref{eq:app:d2J_1}) depends explicitly on mass, through $\delta \phi$ while the dynamic condition (\ref{eq:app:dyn_eig}) does not.
	
\end{document}